# Unraveling the Origin of Mysterious Luminescence peak at 3.45 eV in GaN Nanowires

Swagata Bhunia,[1,2] Soumyadip Chatterjee,[2] Ritam Sarkar,[2] Dhiman Nag,[2] Suddhasatta Mahapatra,[1,*] and Apurba Laha[2,*]

1. Department of Physics, Indian Institute of Technology, Bombay, Powai, Mumbai-400076.
2. Department of Electrical Engineering, Indian Institute of Technology, Bombay, Powai, Mumbai-400076.

- Corresponding authors email: suddho@iitb.ac.in, laha@ee.iitb.ac.in

Abstract: The demand for GaN Nanowires (NWs)-based optoelectronic devices has rapidly increased over the past few years due to its superior crystalline quality compare to their planar counterparts. However, NWs-based devices face significant challenges because of number of surface states, basal plane stacking faults and coalescence related defect states. While the origins of most of the defect states have been identified and mitigated using well-established methods, the origins of few defect states remain unknown, and thus their suppression methods have yet to be explored. One such defect state is the 3.45 eV luminescence peak, known as the UX-band. In this report, we have investigated the origin of this peak in NWs grown on a sapphire substrate by using Plasma Assisted Molecular Beam Epitaxy (PAMBE) tool. We have found that the defect states, generated due to oxygen incorporation, especially when the oxygen atoms substitute the Ga atoms, are the primary cause of this band. Actually the excitons are localized at this defect center and the radiative recombination of localized excitons gives the characteristic UX-band. By protecting the NWs from oxygen incorporation through AlN encapsulation process, we have completely suppressed the 3.45 eV peak and proved further that the peak is caused by oxygen induced defect states. Thus we have addressed an issue that persisted over the last three decades, potentially paving the way for efficiency improvements in optoelectronic devices.

In recent years, GaN nanowires (NWs) have emerged as a potential candidate for optoelectronic and high temperature devices, such as laser diodes (LDs),[1] light emitting diodes (LEDs)[2], building block for single photon sources [3] and single GaN NW p-i-n diode.[4] The key property that has drawn significant attention from the research community is the dislocation-free nature of the NWs. Because the strain induced by lattice mismatch is relaxed through the sidewall of the NWs, they remain free of dislocations.[5, 6] The absence of threading dislocations, coupled with a high surface-to-volume ratio, makes NWs optically and structurally superior to their planar counterparts.[7] Consequently, opto-electronic devices fabricated on NWs demonstrate higher internal quantum efficiency and improved light extraction efficiency compared to those based on planar structure. [8, 9] Despite these advantages, the GaN NW based devices face several challenges, primarily related to surface states,[10, 11] basal plane stacking faults (BSFs), [12, 13] and an unknown band (UX) located at an energy level of 3.45 eV.[14] Although, various processes have been developed to passivate the surface states of NWs and eliminate the stacking faults, no effective experimental methods have yet been identified to suppress the UX band. The primary reason is that the origin of this peak remains poorly understood.

Despite nearly three decades since the first successful growth of GaN NWs, no definitive conclusion has been reached regarding the origin of 3.45 eV peak. Several research groups around the world have attempted to uncover the origin of this peak. For example, based on the Ga balling model of NW-growth, E. Calleja et al., reported that the origin of the peak is Ga-interstitials.[14] In contrast, P. Corfdir et al., through their observation of the decay profile of the 3.45 eV peak in time-resolved PL, claimed that the peak is due to donor bound excitons, located at the sidewall of NWs, which they referred to as two-electron-satellites (TES).[15] Later, P. Lefebvre et al. and O. Brandt et al. suggested that the peak arises from point defects located at the NW surface.[16, 17] More recently P. Huang et al. demonstrated that the NW surface microwires are the primary cause of this peak.[18] On the other hand, C. Pfuller et al. and T. Auzelle et al. argued that the peak is due to the recombination of excitons trapped in the prismatic inversion domain boundary (pIDB). [19, 20] It is evident that the exact origin of the UX band remains unclear. Moreover, there are no reports, describing a process for eliminating this peak.

In this report, we have investigated the origin of 3.45 peaks in GaN NWs grown on a sapphire substrate. Our detailed investigation reveals that these peaks arise from surface states of NWs, specifically oxygen induced defect states that form at NW-surface when they are exposed to oxygen environment. The excitons are actually trapped in the defect states that are 33 meV below the conduction band. The radiative recombination of the localized excitons gives the characteristic UX band. We also observed that the encapsulation of NWs with AlN immediately after the growth suppresses the UX band by protecting the NW-surface from oxygen incorporation and inhibiting the formation of oxygen induced defect state. This proves further that the UX band is associated with oxygen induced surface states. Further investigation shows that the intensity of UX band becomes negligible when an AlN layer of thickness ~ 15 nm is deposited.

Experimental:

All samples were grown on sapphire substrates using a RIBER C21 Plasma Assisted Molecular Beam Epitaxy (PAMBE) tool, equipped with solid sources (Ga, Al, and In) and doping sources (Mg and Si). For nitrogen plasma generation, RIBER add-on plasma cell was utilized. In previous reports, the diameters of the NWs were reduced by growing them at higher growth temperature and using Si as the substrate.[16] Higher growth temperatures create Ga vacancies, and the formation energy of these vacancies decreases with increasing Si incorporation. These Ga vacancies have been observed to enhance the incorporation of oxygen into NWs.[21] Since Si incorporation occurs at higher growth temperatures, use of Si as substrate for GaN NWs growth is not a suitable solution for studies focused on defect states. On the other hand, exposure to oxygen or air increases the density of surface states. Therefore, to accurately investigate the origin of NWs' defect states, encapsulating them with another III-Nitride material post-growth is preferable. To investigate this, two series of NW samples were prepared on sapphire substrates: one without encapsulation layer (A-series) and one with an encapsulation layer (E-series). The A-series consists of GaN NWs grown at three different substrate temperatures: 860 °C, 880 °C, and 900 °C, labeled as samples A1, A2, and A3, respectively. The E-series involves GaN NWs grown at 880 °C, followed by encapsulation of AlN (at 880 °C growth temperature) of varying thickness immediately after the growth. The AlN thicknesses were as 5.8 nm, 9.7 nm, and 15.3 nm, corresponding to sample E1, E2, and E3, respectively. Prior to the GaN NW growth, the sapphire substrates were degassed at 890 °C for 1 hr, then nitridated at 800 °C for 45 min. Following nitridation, the AlN was deposited for 8 min. Subsequently, the substrate temperature was raised to 880 °C, and the shutters of Ga and nitrogen plasma cells were opened for 3.5 hrs to grow the GaN NWs. For E-series samples, AlN encapsulation layers were deposited immediately after NW growth, with deposition times of 3 min, 7 min and 15 min for samples

E1, E2 and E3, respectively. During GaN NW and AlN layer growth, the nitrogen plasma power and flow rate were maintained at 450 Watt and 3.0 sccm, respectively. Whereas, during the growth of AlN layer, these parameters were fixed at 260 Watt and 0.65 sccm. Throughout all process, the beam equivalent pressure (B.E.P) of Ga and Al cells were maintained at $2.1 \times 10^{-7}$ Torr and $8.2 \times 10^{-8}$ Torr, respectively.

Results and Discussion:

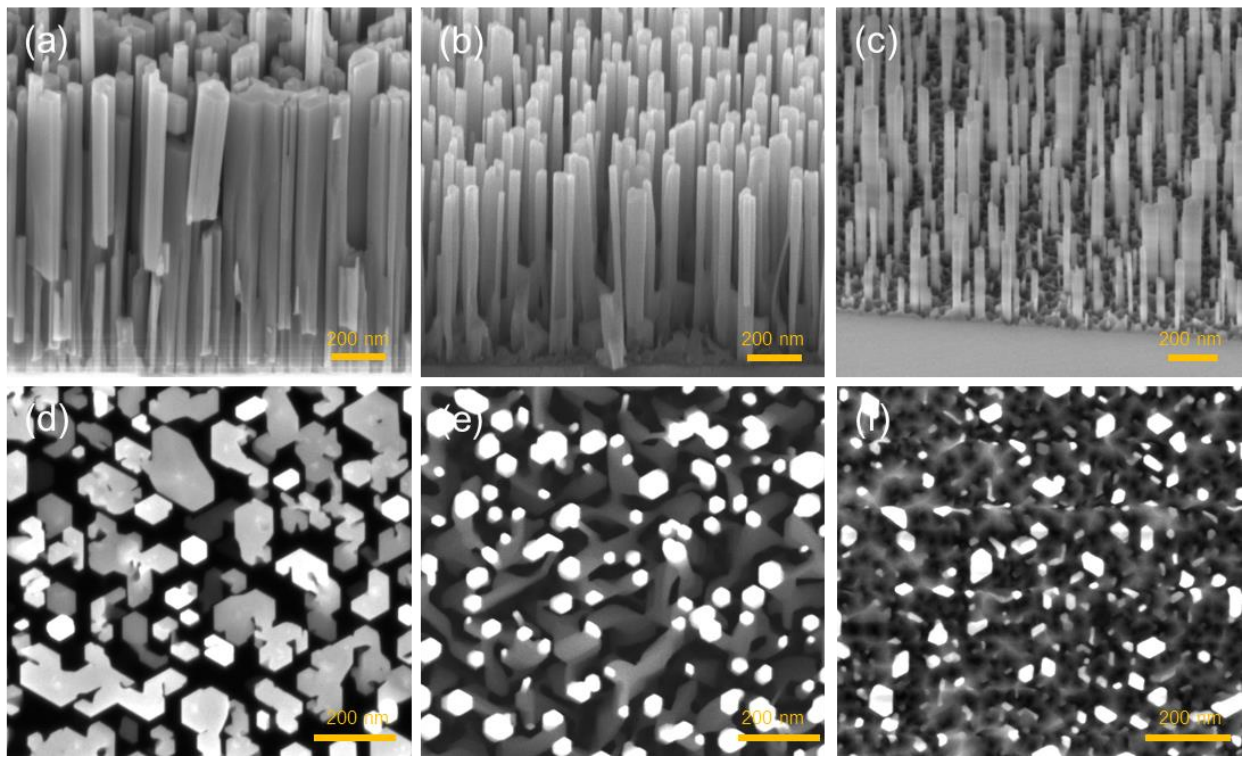

**Figure 1:** (a-c) show SEM 45°-tilted view images of NWs of sample A1, sample A2 and sample A3 while (d-f) represent the SEM-top view images of the same samples. From sample A1 to sample A3, a significant reduction in lengths and diameters of the NWs are observed.

Figure 1 (a-c) show the SEM 45°-tilted view images of NWs grown at substrate temperatures 860 °C (A1), 880 °C (A2), and 900 °C (A3), respectively. The measured lengths of the respective NWs are 1014 nm, 653 nm and 360 nm. From these images, it is clear that as the growth temperature increases, the lengths of NWs decrease. On the other hand, the SEM-top view images, shown in fig. 1(d-e), reveal that the diameter of the NWs also decreases with increasing growth temperature. The average diameters of NWs for samples A1, A2 and A3 are measured to be 95 nm, 44 nm and 30 nm, respectively, indicating that higher growth temperatures, results in NWs with reduced diameters. Considering the NWs as cylindrical in shape, their average surface areas were estimated also. For samples A1, A2 and A3, the average NW surface areas within a $2.7 \times 10^{-8}\ cm^2$ region were evaluated as $3.4 \times 10^{-7}\ cm^2$, $2.5 \times 10^{-7}\ cm^2$ and $1.2 \times 10^{-7}\ cm^2$, respectively. This trend reveals that the surface area of the NW decreases with an increase in growth temperature, this is because with the increase of growth temperature the diameter as well as the length of the NWs decreases. Additionally, since the NWs were grown on sapphire substrate, there is no possibility of Si incorporation to these NWs, eliminating the effect of Si on Ga vacancy related defect states.

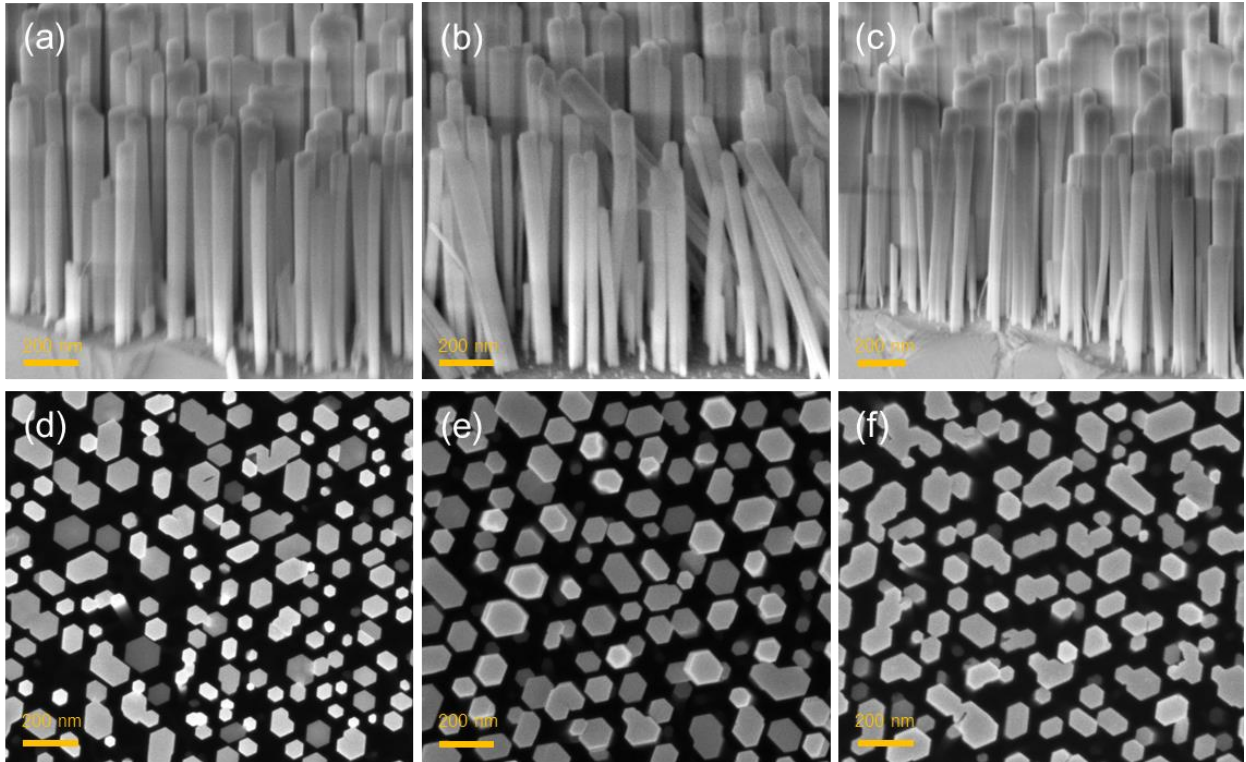

**Figure 2:** (a-c) represent the SEM 45°-tilted view images of NWs of sample E1, E2 and E3 while (d-f) represent the SEM-top view images of the same samples.

For the E-series samples, the SEM 45°-tilted view and top-view images of NWs are shown in fig. 2(a-f). For samples E1, E2 and E3, the measured lengths (diameters) are 811 nm (66 nm), 807 nm (61 nm) and 812 nm (64 nm), respectively. The SEM images clearly indicate that the average length and diameter of all the AlN-capped NWs remain the same. Even though the AlN encapsulating layer was deposited for up to 15 min, no significant change in NWs' diameter was observed.

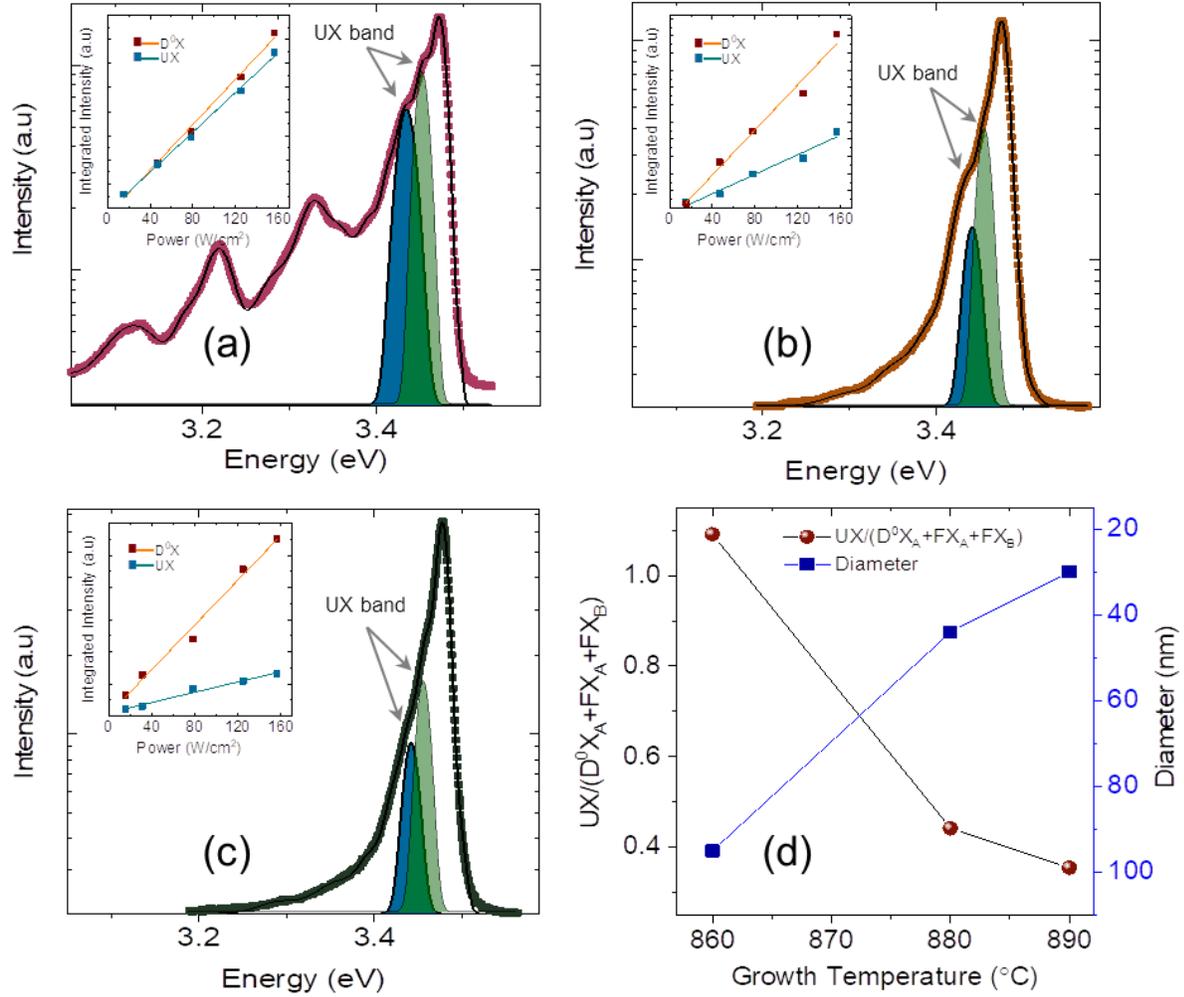

**Figure 3:** (a-c) 10 K PL spectra of sample A1, sample A2 and sample A3. The Gaussian fits of the unknown bands are represented in blue and green color. The insets of (a-c) illustrate the power-dependent PL of $D^0X$ and UX for the respective samples, which were measured at 10 K. (d) Variation of $\frac{(UX)}{(D^0X_A+FX_A+FX_B)}$ ratio with growth temperature and NW-diameter.

The PL spectra of samples A1, A2 and A3 are shown in fig. 3(a-c). The peak with the highest intensity is attributed to the donor bound exciton ($D^0X$), while the two peaks located in close proximity to the $D^0X$ peak correspond to the UX band. Additional peaks are associated with LO-phonon replicas and coalescence-related defect states in the NWs. To determine the exact positions of these peaks, the spectra were fitted with Gaussian functions (with 99% confidence bound). The $D^0X$ and UX band were identified at 3.472 eV, 3.445 eV and 3.452 eV, respectively. For NWs grown on Si substrate, the $D^0X$ level are observed at the same energy level but the UX bands were shifted to slightly higher energy level which are 3.452 eV and 3.458 eV. [14] This shifting may be due to use of different substrate and growth conditions. The fitted curves (shown in fig. 3(a-c)) clearly demonstrate that the band comprises of two distinct peaks, located at 3.445 eV and 3.452 eV, respectively. The PL intensity of these peaks (represented by the total area under the two curves) decreases from sample A1 to sample A3. In earlier section, the FESEM images of NWs reveal a reduction in diameter when comparing NWs in sample A1, sample A2 and sample A3. This suggests that the PL intensity of GaN NWs' UX band become less prominent as the diameter of NWs decreases. This trend can be directly observed from fig. 3(d), where the ratio of UX-band-intensity to total exciton-intensity $\frac{(UX)}{(D^0X_A + FX_A + FX_B)}$ is compared for NWs grown at different growth temperatures and with varying diameters. As the total surface area of NWs reduces with diameter shrinkage, it can be inferred from this observation that the UX band of NWs is related to the surface. Also, this suggests that the UX band is not associated with donor bound excitons located at the surface. If the surface donor bound excitons were responsible for the origin of UX band, the intensity would increase rather than decrease, as observed in earlier reports.[16] To further confirm this, power dependent PL measurements were performed on each sample, as shown in the inset of fig. 3(a-c). The power dependent PL of sample A1 (inset of fig. 3(a)) exhibits a linear increase in the intensity of both $D^0X$ and UX peaks with increasing laser power. A similar trend is also observed in case of samples A2 and A3. However, the variation in the UX band intensity relative to the $D^0X$ peak intensity differs among the samples. Compared to sample A1, sample A2 exhibits a smaller variation in UX band intensity (inset of fig. 3(b)), and this variation becomes even less pronounced in sample A3 (inset of fig. 3(c)). This observation further supports the conclusion that the UX band is related to surface states not to $D^0X$. Since the ratio $\frac{(UX)}{(D^0X_A+FX_A+FX_B)}$ is lower for NWs of lengths 360 nm than 1014 and 653 nm, the UX band is not associated with pIDBs. In this case, the laser power was raised to its maximum value, 156 W/cm$^2$; however, no intensity saturation was observed for either peak ($D^0X$ or UX). Interestingly, these findings are in contrast with earlier reports, where a pronounced increase in UX band-intensity was observed with decreasing NW diameter. [23] This discrepancy is likely due to the incorporation of Si atoms into the NWs at higher temperatures, leading to an increase of Ga vacancies, which may have induced the formation of the UX band. [24] It can also be noticed from the reference [16] that the density of NWs grown at 810 °C is significantly lower, while the density of NWs grown for the current study at 900 °C is higher. This implies that compare to sapphire in case of Si substrate, the effective-substrate temperature is higher, consequently Ga desorption is higher which might have created more number of Ga vacancies.

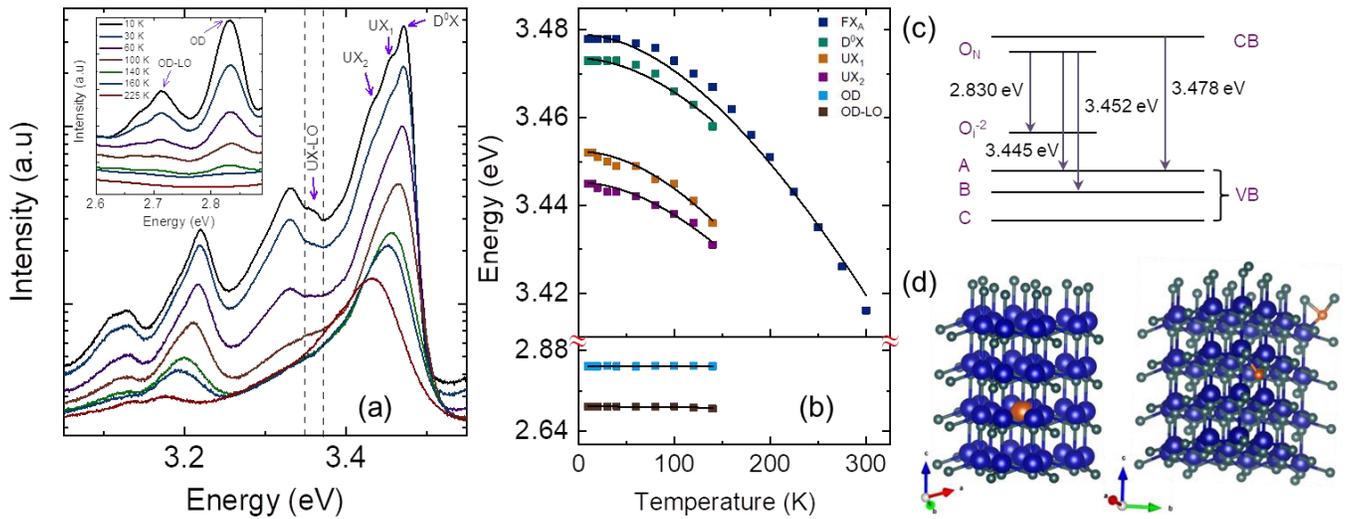

**Figure 4:** (a) represents the temperature dependent PL spectra of sample A1 and in the inset the same spectra taken at lower energy range is shown. (b) shows the variation of energy with temperature of $FX_A$, $D^0X$, $UX_1$, $UX_2$, OD and OD-LO. In (c), the position of oxygen induced defect states within the band structure, along with the transition energies related to these defect states is schematically illustrated. (d) exhibits the position of an oxygen atom in the GaN lattice structure, when it creates substitutional ($O_N$) defect state (left) and interstitial defect state ($O_i$) (right) (simulated using the VESTA software [22]).

In the preceding section it has been demonstrated that the UX band is associated with NW surface states. However, the specific elements contributing to this process remain unidentified. To further investigate, temperature-dependent PL was carried out on all samples. Fig. 4(a) shows the temperature-dependent PL spectra of sample A1, where the peaks (extracted from Gaussian fitting) positioned at 3.478 eV, 3.473 eV, 3.445 eV, 3.452 eV, 3.349-3.372 eV are free-exciton ($FX_A$), donor bound exciton ($D^0X$), $UX_1$, $UX_2$ (two UX lines) and their LO-phonon replicas (UX-LO) (indicated by the arrows). In the inset of fig. 4(a), two peaks are clearly observed at 2.830 eV and 2.714 eV. The peak at 2.830 eV corresponds to point defects created by incorporated oxygen at the NW-surface while the other peak is a result of LO-phonon coupling with this defect related emission. [25] It is interesting to note that energy of UX lines decay with the increase of temperature and this trend is similar to the decay of $FX_A$ and $D^0X$ lines as shown in fig. 4(b). Moreover, like the $FX_A$ and $D^0X$, both UX lines follow the Varshni's equation for temperature-dependent band gap variation (represented by black colored curve), [26] which is $E(T) = E(0) - \frac{\alpha T^2}{(\beta+T)}$ (where $E(T)$ and $E(0)$ are the band gap at temperature T and 0 K, and α and β are the Varshni's thermal coefficients). While the $FX_A$ and $D^0X$ lines follow the Varshni's equation, the oxygen induced state at 2.830 eV (OD) and their LO phonon replica at 2.714 eV (OD-LO) do not emulate it since these states are defect states. Surprisingly, the PL intensity of these defect states vanishes at 140 K, which is the same temperature at which the UX lines disappear also. This implies that UX band has a direct correlation with the oxygen induced defect states. Actually, whenever the oxygen atoms incorporate into GaN NWs, it creates two kinds of defect states at the NW-surface, which are interstitial ($O_i$) and substitutional ($O_N$) defect states. The crystal structure of GaN for those two configurations has been simulated using VESTA software and represented in fig. 4(c). Previous studies reveal when the configuration is fig. 4(c)-left that implies the $O_N$ state then it exhibits a value of 0.033 eV below the conduction band, whereas for another configuration

fig-4(c)-right, which is $O_i$ state that shifts 0.615 eV above the valence band. [27] The energy difference between these two states is 2.830 eV, which is same as the transition energy observed in PL spectra. In the case of GaN NWs, these defect states act as trapping centers for excitons. Since the oxygen atom has a higher electronegativity than nitrogen atom, it more effectively captures electrons, when carriers are generated through laser excitation, and becomes negatively charged. The negatively charged oxygen atom then attracts a photo-generated hole in GaN by long-range coulomb interaction, binding the exciton to that defect state. While an electron in $O_N$ state and the corresponding hole remains at valence band A, their radiative recombination results in an emission line at 3.445 eV ($UX_1$ line). On the other hand, if the hole remains at valence band B instead of A, then the 3.452 eV line ($UX_2$ line) is observed. This implies that the radiative recombination of the trapped exctions causes the UX-band. Additionally, the phonon-replicas of UX band (UX-LO) are observed at 3.349-3.372 eV, which is ~ 90 meV less than the UX band energy level. The exact position of UX-LO peak for individual UX line was very difficult to extract. Thus the observations provide direct evidence that UX bands are generated due to the oxygen induced defect states located at the surface of NWs.

Since the AlN layer protects the NWs surface from environmental exposure, thereby reducing the incorporation of oxygen through the NWs-surface, [28] we capped the NWs with the AlN to investigate whether the encapsulation suppresses the UX band. The 10 K PL spectra of encapsulated NWs are shown in fig. 5(a-c). For NWs in sample E1, the UX band is still present but the $\frac{(UX)}{(D^0X_A+FX_A+FX_B)}$ ratio decreases to 0.98, compared to 1.12 in the uncapped sample A1. This ratio further decreases to 0.52 in sample E2 and become almost negligible at 0.10 in sample E3. This implies that by reducing the oxygen incorporation, the AlN layer suppresses the UX band formation. Additionally, the power-dependent PL measurements for samples E1 and E2 show that the intensity of $D^0X$ and UX peaks increases linearly with power, similar to the behavior observed for the A series samples. However, in the case of sample E3, while a linear increase in the $D^0X$ was observed, the UX-band intensity remains nearly constant. It is important to note that despite the NWs having the same diameter and length, the UX band intensity decreases and remains constant in sample E3, which is attributed to the increase thickness of AlN encapsulation layer. This observation suggests that the AlN encapsulation layer effectively suppresses the formation of such defect states, with this suppression becoming more effective as the AlN thickness increases. Additionally, the PL spectra of samples E2 and E3 show negligible peak intensities at 2.830 eV and 2.714 eV, whereas, the uncapped NWs (sample A1) exhibits prominent intensity at the same energy levels (fig. 5(d)). In the case of AlN capped NWs (sample E2 and sample E3), the absence of these PL peak and the reduction of UX band provide a direct proof that the UX band originates from oxygen induced surface states. It is noteworthy that in spite of capping the NWs with AlN, sample E1 exhibits a peak at 2.830 eV.

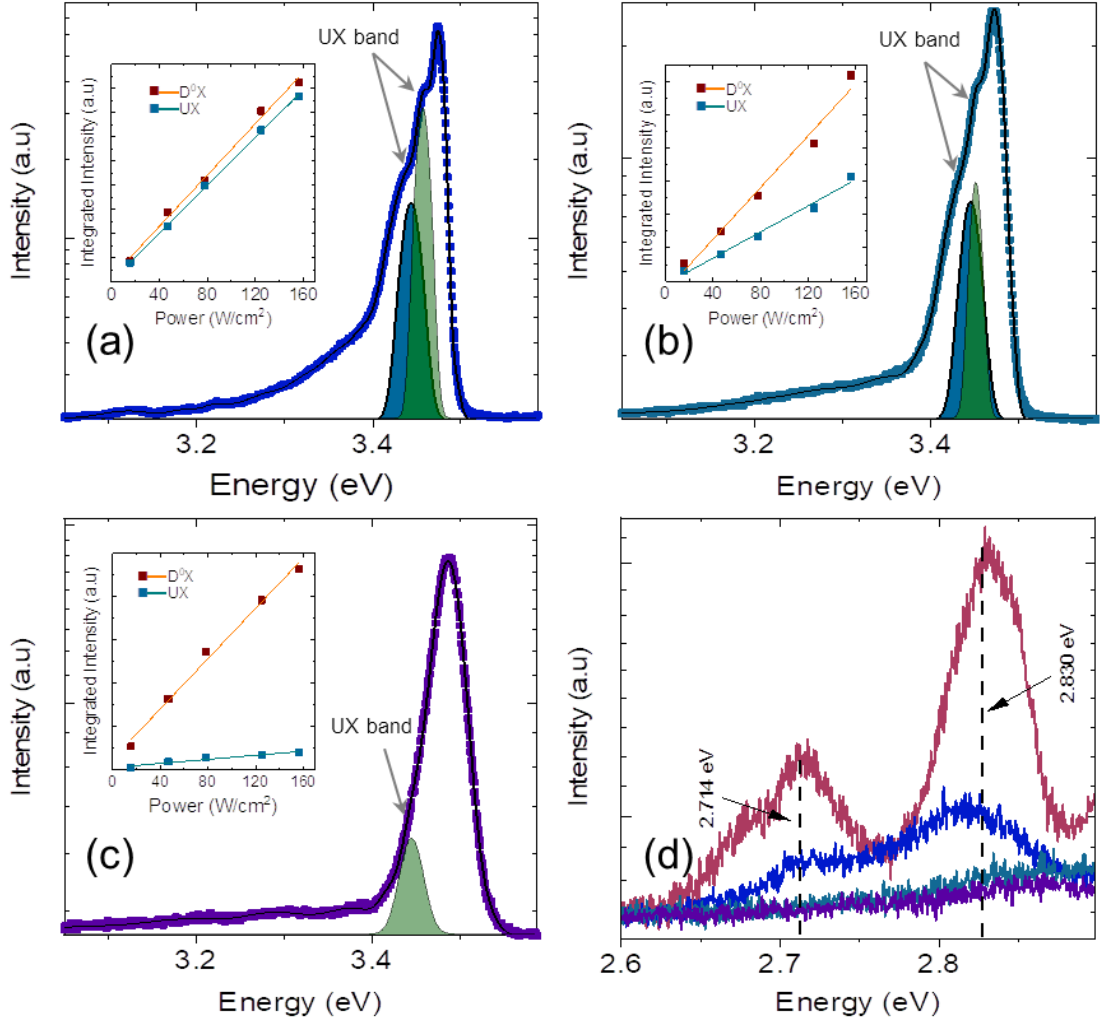

**Figure 5:** (a-c) 10 K PL spectra of sample E1, sample E2 and sample E3. The UX band peaks are represented with blue and green colored Gaussian curves. The variation of $D^0X$ and UX band intensity with laser power measured at 10 K are shown in the insets of (a)-(c). The PL peaks related to oxygen induced defect states are displayed in (d).

To investigate this further, high resolution transmission microscopy (HRTEM) was conducted on all AlN capped samples, as shown in fig. 6(a-c). The TEM images reveal that AlN grows epitaxially both on the top surface and along the sidewalls of NWs. However, higher magnification images (insets highlighted by orange rectangular boxes) demonstrate that the thickness of the top AlN is greater than that of the sidewall AlN layer. For samples E1 (fig. 6(a)), E2 (fig. 6(b)) and E3 (fig. 6(c)), the measured thicknesses of the top AlN-layers are 5.8 nm, 9.7 nm and 15.3 nm, respectively, whereas corresponding sidewall-AlN thicknesses are 1.8 nm, 2.3 nm and 5.4 nm. These observations clearly indicate that the top AlN-layer is thicker than the sidewall-AlN layer, and both thicknesses increase with longer AlN growth time. This suggests that the growth rate of top-AlN is higher than that of sidewall-AlN. However, the higher

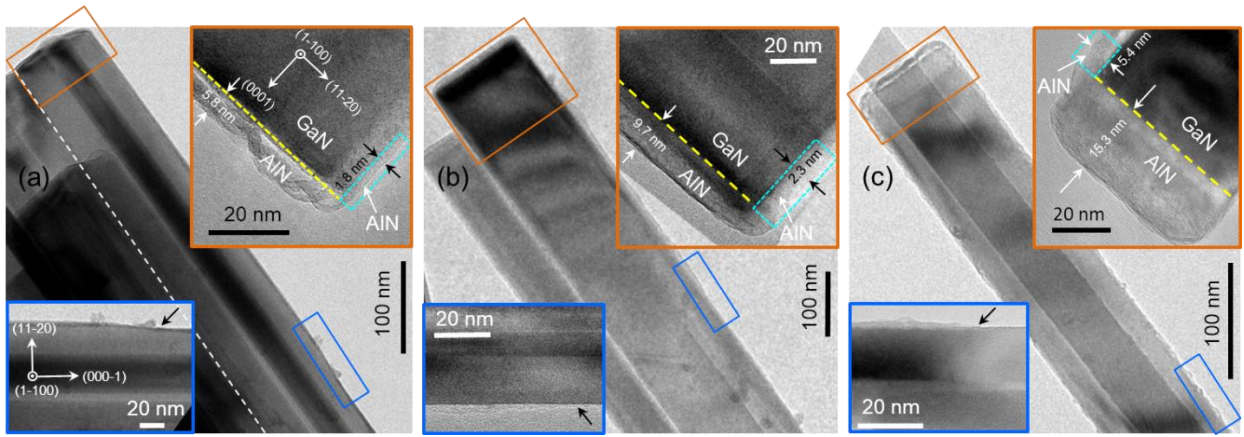

**Figure 6:** (a-c) HRTEM images of NWs for sample E1, E2 and E3, respectively. The insets (orange colored boxes) display higher magnification images of the top portion of the NWs. For samples E1, E2 and E3, the thickness of top (sidewall) AlN layers are 5.8 nm (1.8 nm), 9.7 nm (2.3 nm) and 15.3 nm (5.4

nm), respectively. The higher resolution images, captured at the NWs' sidewall (blue colored boxes) reveal that the dripping down of AlN stops after a certain length along the NWs, as shown by the black arrows. The length is dependent on the AlN thickness.

resolution TEM images (marked with blue-colored rectangular boxes) reveal that the AlN-layer only forms up to a certain length of NWs and does not cover their entire length. For samples E1, E2 and E3, the fractions of the NW lengths that AlN cover are 0.42, 0.57, and 0.73, respectively, as shown in the blue colored rectangular boxes. This entails that the AlN coverage increases with longer AlN growth time. Since the AlN was deposited for a maximum of 15 minutes for sample E3, it shows the highest coverage. In contrast, sample E1, which had a shorter deposition time, shows a lower AlN coverage (482 nm of 1143 nm NW). This incomplete coverage leaves a substantial portion of the NWs exposed, allowing oxygen to incorporate into these regions and create defect states. As a consequence, the PL peak related to oxygen-induced-defect-states is observed in sample E1 (fig. 5(d)).

The previous measurements confirm that the origin of UX band is the oxygen induced defect states. However, the chemical bonding of oxygen atoms and how much of it present in NWs are unknown. For detailed investigation, the XPS measurement was performed for the without AlN-capped and AlN-capped NWs. The spectra recorded for Ga 3d for samples A1, E1 and E3 are shown in fig. 7(a-c). To find out different chemical compounds from the spectra, they were deconvoluted by fitting with Gaussian curves. For all the samples, three dominant peaks are observed at binding energy 19.6 eV, 20.2 eV and 21.1 eV, which are corresponding to Ga-$N_{3d5/2}$, Ga-$N_{3d3/2}$ and Ga-O bonds. The other peaks at 17.4 eV, 18.6 eV and 23.1 eV are due to N 2s, Ga-Ga and Al-O bonds. For sample A1, sample E1 and sample E3, the ratio of Ga-O and Ga-N bonds are found to be 0.187, 0.066 and 0.046, respectively.

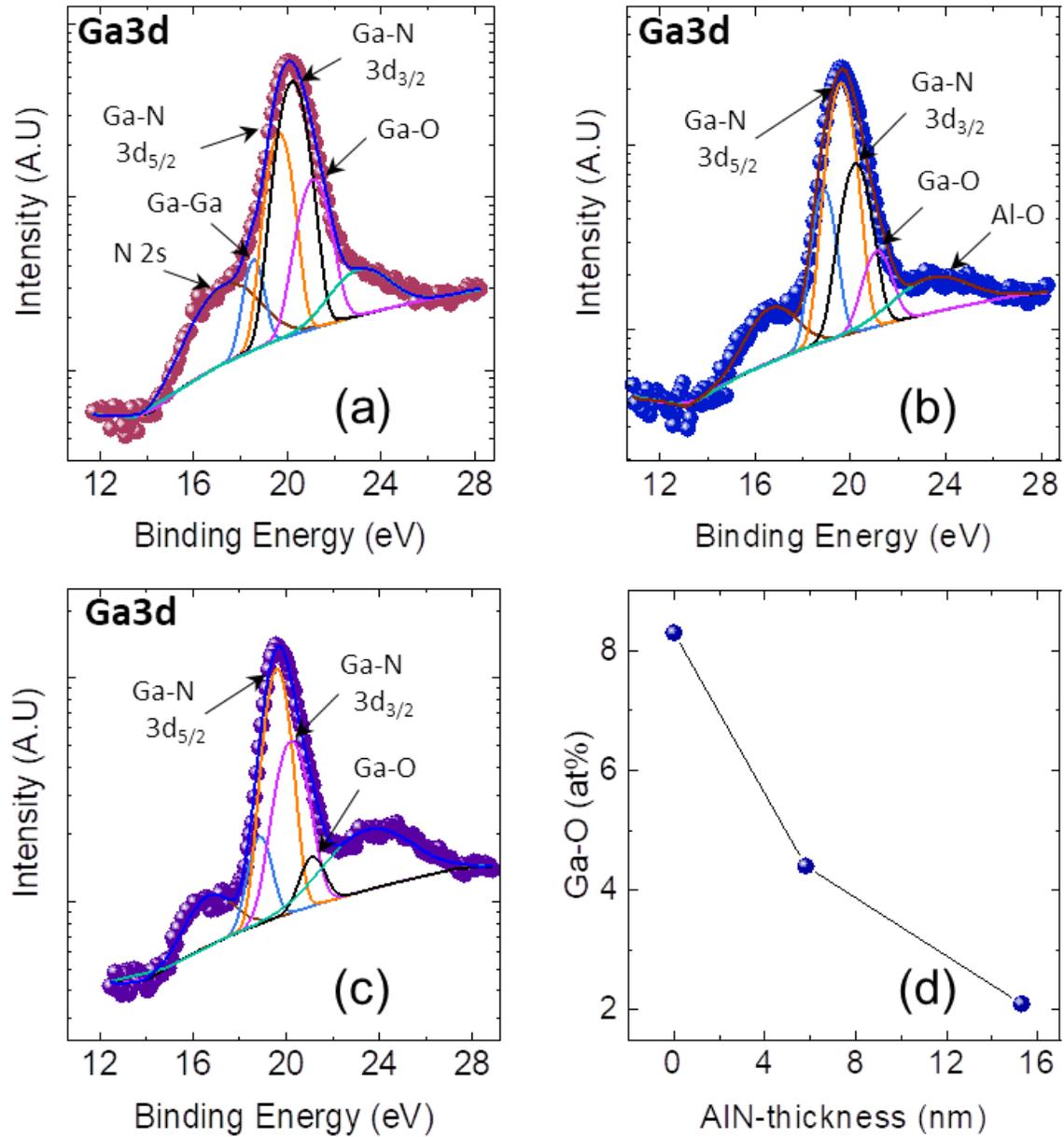

**Figure 7:** (a-c) represent Ga3d XPS spectra for sample A1, sample E1 and sample E3, respectively. The peaks corresponding to individual bonds, identified through Gaussian curve fitting, are indicated by arrows. (d) illustrates the variation of atomic percentage of Ga-O bonds with the AlN thickness.

For sample A1, sample E1 and sample E3, the ratio of Ga-O and Ga-N bonds are found to be 0.187, 0.066 and 0.046, respectively. This indicates that the Ga-O bonds become lesser in case of AlN-capped NWs than the uncapped one. To estimate the concentration of Ga-O bonds, the ratio of area under the curves and the relative sensitivity factor (RSF) of individual element taken from the CasaXPS software database was extracted. The atomic percentage of Ga-O bonds is found to be 8.3 for sample A1; however, in case of sample E1 it reduces to 4.4 and became 2.1 for sample E3 (fig. 7(d)). This observation shows that oxygen is present in the GaN NWs surface and also supports the fact that the encapsulation of GaN NWs with AlN reduces the oxygen incorporation. Moreover, it is noticed that a thicker AlN layer reduces the oxygen incorporation significantly.

Conclusions:

In conclusion, we have identified the origin of the UX band, which appears at 3.445 eV and 3.452 eV due to oxygen-induced surface states. The incorporation of oxygen creates defect states at 0.033 eV below the conduction band where free excitons become trapped. The transition from this state to the valence band A and valence band B, corresponding to the 3.445 eV and 3.452 eV transitions. Furthermore, in GaN NWs encapsulated with AlN, the absence of both the UX band and the oxygen induced defect peaks further confirms that the oxygen incorporation is the primary cause of this band. Thus we have resolved a mystery that was there over the past three decades and demonstrated a method to mitigate the UX band. Thus, this finding will be crucial towards efficiency enhancement of the NWs-based optoelectronic devices.

Acknowledgement:

All authors gratefully acknowledge the financial support from the Science and Engineering Research Board (SERB; project no. CRG/2018/001405) of the Department of Science and Technology (DST), the Quantum Information Technologies with superconducting devices and Quantum Dots (Project no- RD/0120-DSTIC01-001), and the Nano-electronics Network for Research and Applications (NNetRA) (Project Code No. RD/0520-IRNTRS0-001) of the Govt. of India (GoI). The authors also thank the IIT Bombay Nanofabrication Facility (IITBNF) for providing technical support in the execution of this project.